\documentclass[showpacs,reprint,amsmath,amssymb,superscriptaddress,aps,prl]{revtex4-1}
\usepackage[percent]{overpic}
\usepackage{amsmath,amssymb}
\usepackage{graphicx}
\usepackage{float}
\usepackage{xcolor}
\begin{document}

\title{Measurement of complex supercontinuum light pulses using time domain ptychography}

\author{Alexander M. Heidt}\email{Corresponding author: alexander.heidt@iap.unibe.ch}
\affiliation{Institute of Applied Physics, University of Bern, Sidlerstrasse 5, 3012 Bern, Switzerland}
\author{Dirk-Mathys Spangenberg}
\affiliation{Laser Research Institute, Stellenbosch University, Private Bag X1, 7602 Matieland, South Africa}
\author{Michael Br\"ugmann}
\affiliation{Institute of Applied Physics, University of Bern, Sidlerstrasse 5, 3012 Bern, Switzerland}
\author{Erich G. Rohwer}
\affiliation{Laser Research Institute, Stellenbosch University, Private Bag X1, 7602 Matieland, South Africa}
\author{Thomas Feurer}
\affiliation{Institute of Applied Physics, University of Bern, Sidlerstrasse 5, 3012 Bern, Switzerland}

\date{\today}


\begin{abstract}
We demonstrate that time-domain ptychography, a recently introduced ultrafast pulse reconstruction modality, has properties ideally suited for the temporal characterization of complex light pulses with large time-bandwidth products as it achieves temporal resolution on the scale of a single optical cycle using long probe pulses, low sampling rates, and an extremely fast and robust algorithm. In comparison to existing techniques, ptychography minimizes the data to be recorded and processed, and drastically reduces the computational time of the reconstruction. Experimentally we measure the temporal waveform of an octave-spanning, 3.5~ps long supercontinuum pulse generated in photonic crystal fiber, resolving features as short as 5.7~fs with sub-fs resolution and 30~dB dynamic range using 100~fs probe pulses and similarly large delay steps.
\end{abstract}

\maketitle

Supercontinuum (SC) generation has become a scientific and commercial success story, in particular driven by photonic crystal fiber (PCF) technology.
Tailored fiber-based SC sources delivering spectral bandwidths of one octave and more are now routinely applied in bio-photonic imaging, microscopy, frequency metrology, and nonlinear pulse compression, to name only a few \cite{Alfano2016}. Despite such wide-spread use, the full characterization of such SC pulses remains extremely challenging
as their temporal waveforms are amongst the most complex to be found in ultrafast optics. 
Their envelope is typically several picoseconds long, but contains fine structure on the scale of just a few femtoseconds. Combined with their ultrabroad and equally complex sprectrum they easily reach time-bandwidth products (TBP) in the order of $\sim 100-1000$ \cite{Dudley2006}. 

Since no known detector is fast enough to measure such pulses directly, several schemes for the indirect reconstruction of temporal intensity and phase have been devised, of which cross-correlation frequency resolved optical gating (XFROG)  in various implementations has been most successful in adequately measuring complex SC pulses \cite{Gu2002, Dudley2002c, Liu2011b}. It is based on recording the spectrogram of the unknown object pulse cross-correlated via a nonlinear interaction with a series of known time-delayed gate pulses and iteratively reconstructing the electric field using a derivative of the generalized projections algorithm (GPA) \cite{Kane1999}. For the accurate reconstruction of pulses with large TBP  high sensitivity, large ranges, and fine resolution in both temporal and spectral domains have to be achieved simultaneously, which in the case of XFROG entails the recording of a large number of time-gated spectra using short gate pulses. For such complex pulses the GPA demands the interpolation of the measured spectrogram on a time-frequency grid with a number of points in the order of $2^{24}$ or more \cite{Gu2002, Dudley2002c}, all of which makes the experimental measurement and the reconstruction of the pulse a rather cumbersome and computationally intensive task.

Recently time-domain ptychography was introduced as a new modality for ultrafast pulse characterization \cite{Spangenberg2015, Spangenberg2015b, Spangenberg2016}, with an application range even extending down to the attosecond regime \cite{Lucchini2015}. Originating from lensless spatial imaging, ptychography is based on a powerful phase retrieval algorithm, the ptychographic iterative engine (PIE) \cite{Faulkner2005}. While the experimental realization is similar to XFROG in the sense that the cross-correlation spectrogram is recorded, the data processing and the reconstructing algorithm are conceptually very different and offer several key advantages, which become especially relevant in the context of measuring ultra-complex SC pulses: 1) The temporal resolution is independent of both the duration of the gate pulse and the time delay increment. In fact, the gate pulse should be much longer than the object pulse (or longer than the relevant temporal features contained therein). 2) Typically only a few spectra have to be recorded since long gate pulses and large delay steps can be used. 3) The computational grid size required by PIE is orders of magnitudes smaller than the one used in the GPA without compromising temporal resolution, resulting in fast processing and convergence. Here we demonstrate through simulations and experiments that these properties make time-domain ptychography an excellent choice for the efficient and accurate measurement of complex pulses with large TBP.

\begin{figure}[t]
\centering
\includegraphics[width=\linewidth]{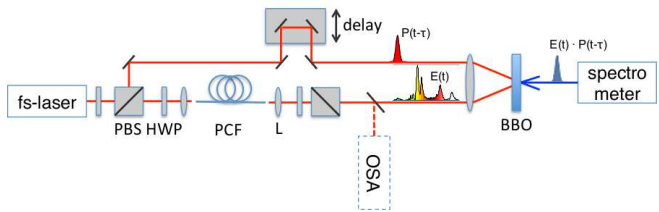}
\caption{Schematic experimental setup. PBS: polarizing beam splitter; HWP: half-wave plate; L: lens; PCF: photonic crystal fiber; OSA: optical spectrum analyzer.}
\label{fig:setup}
\end{figure}

Fig.~\ref{fig:setup} shows a schematic setup for the implementation of time-domain ptychography of SC pulses. A modelocked Ti:Sapphire oscillator generates Fourier-limited pulses with 100~fs full width at half maximum (FWHM) duration centered at 800~nm with 1~W average power at 76~MHz repetition rate. The pulses are split into two copies with variable power ratio using a polarizing beam splitter (PBS) and a half-wave plate (HWP) and serve both as the pump for the SC generation as well as the gate for the recording of the spectrogram. The pump pulse is coupled into the fast axis of a 22.5~cm long polarization maintaining PCF (NKT NL-750 PM) with a zero-dispersion wavelength (ZDW) of 760~nm. The generated SC is collimated and passes through a set of achromatic HWP and PBS that allows to adjust the polarization direction and discard any SC components that might have become depolarized in the PCF. The cross-correlated spectrogram
\begin{equation}
S(\omega,\tau) = \left | \int dt E(t) P(t-\tau) \exp(-i \omega t) \right |^2
\label{eq:spectrogram}
\end{equation}
is then generated by non-collinearly overlapping the SC field $E(t)$ with the time-delayed gate $P(t-\tau)$ in a $30~\mu$m thick nonlinear BBO crystal and recording the spectrum of the sum-frequency generation (SFG) signal for different time delays $\tau$. Note that it is also possible to use other forms of nonlinear interaction with different gate functions, such as phase-only gates \cite{Lucchini2015}.

We first perform numerical simulations closely modelled after the experimental setup in order to verify the performance of PIE for pulses with large TBP. We model the SC generation process by numerically solving the generalized nonlinear Schr\"{o}dinger equation (GNLSE) using the dispersive and nonlinear properties provided by the PCF manufacturer \cite{Rieznik2012}. The model has a temporal and angular spectral resolution of $\delta t = 0.85$ fs and $\delta \omega = 0.9$ THz, we assume a pump pulse peak power of 2.5~kW, and also take the collimating, focusing, and polarizing optics after the PCF into account. Since it is well known that the generated SC might suffer from substantial pulse-to-pulse fluctuations if the nonlinear broadening mechanisms are sensitive to shot noise \cite{Gu2002}, we also included quantum noise terms into the GNLSE and calculated the spectrally averaged first order degree of coherence $\langle g_{12} \rangle$ from a set of multiple simulations with random noise seeds \cite{Dudley2006}. For our conditions $\langle g_{12} \rangle > 0.98$, indicating that the spectrogram averaged over a large number of shots does not differ significantly from a single shot measurement. Hence we implement our experiment in a simple multi-shot setup assuming an identical pulse is available at every shot. Note, however, that a single-shot scheme for measuring complex spectrograms has recently been devised for a variant of XFROG \cite{Wong2014}, which could easily be adapted for ptychography should this become necessary for measuring pulses with larger fluctuations.

We use the result of the GNLSE simulation to calculate the spectrogram from Eq.~(\ref{eq:spectrogram}) and pass it to the PIE reconstruction algorithm, which works on two essentially independent grids. The object and gate pulse are sampled on a temporal grid with $M$ samples equally spaced by $\delta t$, which is determined by the resolution and the range of the spectrometer used such that $\delta\omega \delta t = 1/M$. If the spectrogram contains the full spectral range of the pulse, zero padding can be used to increase $M$ and consequently adjust the desired temporal resolution of the reconstruction. We choose $M=2^{13}$ such that $\delta t = 0.85$ fs, both for simulations and experiments. The second grid is that of time delays and consists of $N$ samples equally spaced by $\delta \tau$. Ptychographic reconstruction works well as long as $\delta \tau$ is smaller than the FWHM width $T_p$ of the gate pulse, i.e. for a sampling rate $R = T_p/\delta \tau > 1$. The resulting redundancy is the basis for the accuracy and robustness of the PIE algorithm. Thus $\delta \tau$ can be chosen freely within the constraint $R>1$. $N$ is typically orders of magnitude smaller than $M$, for example in this paper $N$ ranges between 3 and 103, depending on $R$ and the desired delay range to be reconstructed. Hence PIE requires $N$ cross-correlated spectra recorded at different time delays $\tau_n$ between the object and gate pulse, which are combined in the spectrogram $S(\omega,\tau)$ sampled on a $MxN$ grid. No further interpolation is required. In contrast, GPA based schemes work with a spectrogram that has to fulfill the discrete Fourier transform relations and hence needs to be interpolated on the delay axis to a $M \times M$ grid \cite{Kane1999}. For achieving the same temporal resolution as we do with PIE, an enormous grid size of $2^{13} \times 2^{13}$ would have to be processed, independent of the actual delay range of interest. PIE therefore operates with 100 - 3000 times smaller grid size than the GPA. With typically hundreds of iterations needed before convergence is reached, PIE saves computational time in the order of hours compared to GPA on a standard modern PC. This impressively demonstrates the power of ptychographic reconstruction, which becomes especially apparent when large temporal and spectral ranges have to be combined with fine resolution. 

\begin{figure}[h]
\centering
\includegraphics[width=\linewidth]{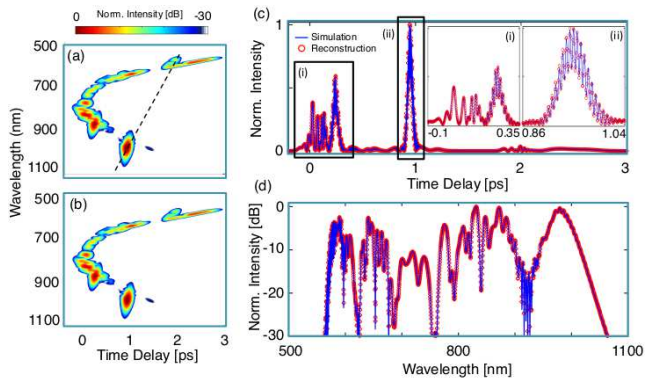}
\caption{Comparison of simulated and reconstructed SC pulse. (a) Simulated and (b) reconstructed spectrogram, (c) temporal intensity and (d) and logarithmic spectrum. The dashed line in (a) connects wavelength components that overlap at the fiber exit.}
\label{fig:simulation}
\end{figure}

The simulated and reconstructed spectrograms ($8192 \times 103$ arrays) with $T_p = 100$ fs and $\delta \tau = 35$ fs (i.e. $R \simeq 3$) are displayed in Figs. ~\ref{fig:simulation}(a) and \ref{fig:simulation}(b) on a logarithmic intensity scale with 30~dB dynamic range. The spectrum spans the range 570 - 1050 nm, the duration of the pulse is about 3.5 ps, and the TBP is > 500. The agreement between simulation and reconstruction is excellent, both are visually indistinguishable. The spectrogram is dominated by the group index profile of the PCF modified by a tilt caused by the dispersive optics in the setup (dashed line in Fig.~\ref{fig:simulation}(a)), and consists of a normal (top) and anomalous (bottom) dispersion arm. The compact objects in the anomalous dispersion arm can be identified as solitons, while the low-level radiation in the top arm are the corresponding dispersive waves. The temporal profile of the pulse is dictated by the overlap of these two dispersive arms and the resulting interference structures, which are particularly sensitive to the relative phase of the overlapping components. Two especially challenging structures for the reconstruction are highlighted in Fig.~\ref{fig:simulation}(c): (i) the densly packed solitons around $\tau = 0$, where small relative phase errors will result in large intensity errors; and (ii) the overlap of the strongest soliton with a dispersive wave at $\tau \approx 950$~fs, which causes an extremely fast temporal beating with 6.5~fs period, equal to the frequency of their spectral separation of about 150~THz. Both structures are perfectly resolved in the reconstruction. Finally, the comparison of the spectra in Fig.~\ref{fig:simulation}(d) shows that the ptychographic reconstruction is accurate even down to the -30~dB level and hence achieves precise phase retrieval even in sections of the pulse where the spectral intensity is low. This is a particularly important feature for SC pulse characterization, because the temporal profile critically depends on a correct relative phase across the whole bandwidth. The conclusions of the visual comparison are confirmed by the root mean square error (rms) between original and reconstructed spectrogram, 
\begin{equation}
\epsilon = \sqrt{\frac{\sum{[S_{\textnormal{}} (\omega,\tau) - S_{\textnormal{rec}}(\omega,\tau)]^2}}{N M}},
\label{eq:rms}
\end{equation}
which we found to be limited only by machine precision and the computational time required to achieve it. Here we aborted the PIE algorithm when $\epsilon=10^{-5}$ was reached after about 300 iterations. 

\begin{figure}[b]
\centering
\includegraphics[width= \linewidth]{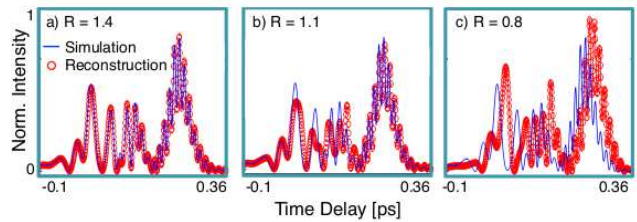}
\caption{Accuracy of the temporal reconstruction in dependence of sampling rate $R$.}
\label{fig:delay}
\end{figure}

Next we investigate the behaviour of PIE for lower sampling rates by keeping $T_p = 100$~fs fixed and increasing $\delta \tau$. Fig.~\ref{fig:delay} shows the results of the temporal intensity reconstruction for $\delta \tau$ equal to 70~fs, 90~fs, and 120~fs, focusing on the soliton group around zero delay for better visibility. For $\delta \tau = 70$~fs ($R=1.4$), the reconstruction is of equal quality ($\epsilon < 10^{-5}$) as the one presented in Fig.~\ref{fig:simulation}, but requires only half the number of spectra and half the grid size ($8192 \times 51$). For $\delta \tau = 90$~fs, $R$ approaches one, further decreasing the required number of spectra, but also the redundancy in the measurement and consequently the rebustness of the reconstruction. Nevertheless, the main features are well recovered, with only small relative phase errors leading to slight discrepancies in the intensity of interference features ($\epsilon = 2.7 \cdot 10^{-4}$). For $\delta \tau = 120$~fs, the measurement is undersampled due to $R<1$, leading to larger errors and we find that $\epsilon$ does not fall below $1.1 \cdot 10^{-3}$. This set of simulations therefore confirms that PIE is able to accurately reconstruct the full complexity of the pulse even for large $\delta \tau$ as long as $R>1$. Note that this condition implies that the number of spectra can be further reduced by using longer gate pulses. Experimentally a limit to this rule might be posed by the required peak power in the nonlinear process that creates the cross-correlated signal.

\begin{figure}[t]
\centering
\includegraphics[width=\linewidth]{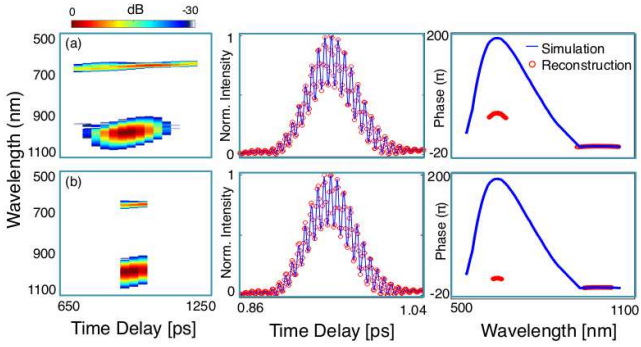}
\caption{Partial reconstruction around $\tau=950$~fs showing spectrogram, temporal intensity and spectral phase for (a) $N=15$ and (b) $N=3$ spectra.}
\label{fig:slices}
\end{figure}

Lastly, we highlight the capability of PIE to do partial reconstructions, a feature that is particularly useful if only short sections of an SC pulse are of interest. As an example, we focus on the fast temporal beating on the envelope of the soliton at $\tau=950$~fs. We again consider $T_p=100$~fs and $\delta \tau = 35$~fs as in Fig.~\ref{fig:simulation}, but reduce the size of the sampled delay window to the region of interest, thus simulating a partial measurement of the pulse. Fig.~\ref{fig:slices} shows the resulting spectrograms and reconstructions of temporal intensity and spectral phase for delay windows of 490~fs and 70~fs, corresponding to $N=15$ and $N=3$ spectra, respectively. In both cases, the temporal waveform is surprisingly well reconstructed. The retrieved absolute spectral phase does not match the simulation, because there is no available information about the intermediary phase between the two isolated spectral components. Intererestingly, however, the relative phase modulo $2\pi$ between the two spectral components is determined accurately, apparent from the correct reconstruction of the phase of the interference pattern in the temporal waveform. Note that in this example PIE correctly resolves features on the time scale of two optical cycles using 100~fs probe pulses and requiring only three spectral measurements.

\begin{figure}[h]
\centering
\includegraphics[width= \linewidth]{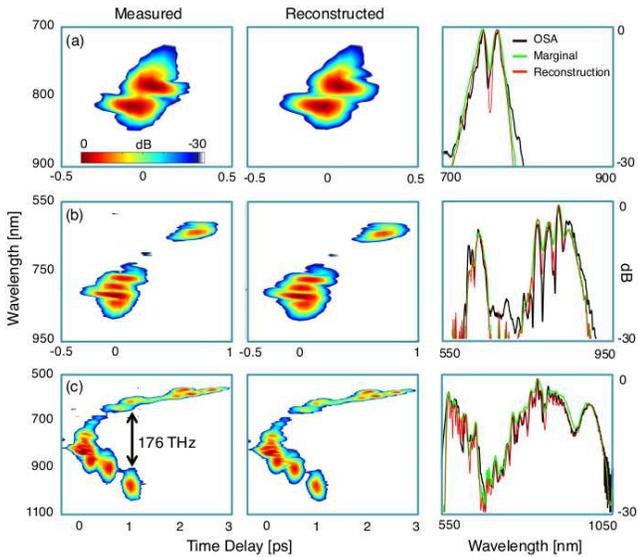}
\caption{Experimental ptychographic measurement of different SC generation stages in 22.5 cm long NL-750 PM PCF, using pump peak powers of 200 W (a), 570 W (b), and 2.5~kW (c). The columns show measured and reconstructed spectrograms, and a comparison of the measured spectral marginal (green), the retrieved spectrum (red), and an independently measured OSA spectrum (black). }
\label{fig:exp}
\end{figure}

We now proceed to the experimental realization using the setup already described above (Fig. \ref{fig:setup}). A practical challenge of this implementation is that the SFG phase-matching bandwidth of the BBO crystal is smaller than the maximum spectral bandwidth of the SC. Fast angle-dithering of the crystal during the measurement has been shown to solve this problem \cite{OShea2000}, but since PIE only requires a small number of spectra we decided to simply repeat the measurement for different crystal angles and obtain the full spectrogram from a superposition after correcting for the phase-matching efficiency versus wavelength. A maximum of three measurements with suitably chosen crystal angles were sufficient to cover the complete SC bandwidth. The comparison of the spectral marginal, i.e. the integral of the measured spectrogram with respect to delay, to an independently measured spectrum ensured the accuracy of the post-processing \cite{DeLong1996}. Standard background correction and Fourier noise filtering were also applied. 

Fig.~\ref{fig:exp} shows measured and reconstructed spectrograms with $T_p=100$~fs after conversion to fundamental wavelengths for pump peak powers between 200~W and 2.5~kW coupled into the PCF. Following the results of the simulations, $\delta \tau = 70$~fs was chosen. The sampling grid size is $8192 \times N$, where $N=[16; 25; 51]$ for Fig.~\ref{fig:exp}(a)-(c), respectively, resulting in a temporal resolution equal to that used previously in the simulations, i.e. $\delta t = 0.85$~fs. We achieved excellent sensitivity allowing us to record all spectrograms with signal-to-noise ratios  (SNR) exceeding 30~dB and enabling us to follow the  development of the SC from low to high pump powers. In Fig.~\ref{fig:exp}(a)-(c) we observe self-phase modulation, soliton fission, and fully developed SC spectrum with a bandwidth of about one octave, respectively. In all cases, the reconstruction is of excellent quality for an experimentally recorded data set, with $\epsilon = 3 \cdot 10^{-3}$ for (a), (b) and $\epsilon = 6 \cdot 10^{-3}$ for the broadest spectrum (c), limited mainly by the SNR. The reconstruction required a computational time of less than 1 minute on a standard laptop computer.

\begin{figure}[htbp]
\centering
\includegraphics[width= \linewidth]{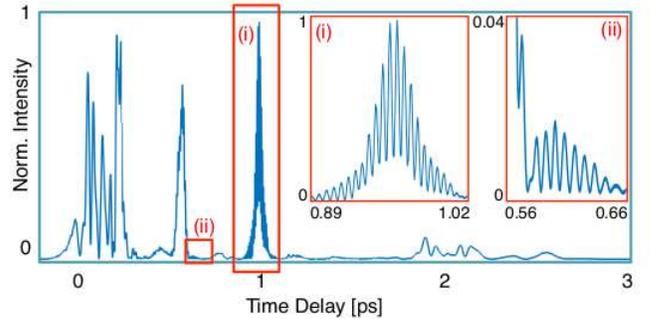}
\caption{Measured temporal intensity of the octave-spanning SC pulse. Insets show fully resolved temporal beatings with period of 5.7 fs (i) and 8.2~fs (ii), the latter with an amplitude as low as $1.5 \cdot 10^{-3}$.}
\label{fig:pulse}
\end{figure}

It is important to confirm that both the measured spectrogram and the reconstruction are correct. In Fig.~\ref{fig:exp} we therefore also show the measured spectral marginal and the retrieved spectrum, and compare both to an independently measured OSA spectrum. In all cases the agreement is very good, confirming the accuracy of both the measurement procedure and the reconstruction. Since there is also a good qualitative agreement between simulated and measured SC spectrogram (compare Figs.~\ref{fig:simulation} (b) and \ref{fig:exp} (c)), we also know that the shortest temporal feature of the fully developed SC pulse should be the beating caused by the overlap of soliton and dispersive wave at $\tau \sim 950$~fs with a beat frequency equal to their spectral separation. From the measured spectrogram we determine that separation to be $\sim 176$~ THz, resulting in an expected oscillation period of about 5.7~fs. Fig.~\ref{fig:pulse} shows the reconstructed temporal waveform with a full range of 3.5~ps, where this expected period is exactly recovered and fully resolved. Interestingly, we also resolve the much weaker temporal beating with a period of 8.2~fs at the edge of the soliton at $\tau \sim 600$~fs, caused by the overlap with a low-level dispersive wave separated by about 120~THz. The amplitude of this oscillation is as low as $1.5 \cdot 10^{-3}$ on the normalized scale. Hence we achieve a dynamic range close to 30~dB in the temporal domain. 

In conclusion we have demonstrated that time-domain ptychography is a powerful technique for the temporal characterization of complex ultrashort pulses with large TBP. It is able to accurately resolve features on the scale of a single optical cycle using 100~fs (or longer) gate pulses and equally large delay steps, minimizes the number of time-gated spectra to be recorded and processed, and is superior to comparable GPA based techniques in terms of memory requirements and computational speed. 

\section*{Acknowledgments}
We gratefully acknowledge financial support from the National Research Foundation, the CSIR NLC, and the NCCR MUST research instrument of the Swiss National Science Foundation.

\end{document}